\documentclass[review,sort&compress]{elsarticle}

\usepackage{amsmath}
\usepackage{booktabs}
\usepackage{graphicx}
\usepackage{hyperref}
\usepackage{lineno}
\usepackage{microtype}
\usepackage{xspace}
\usepackage{siunitx}

\hypersetup{pdfauthor=author}
\newcommand*{\kovats}{Kov\'ats\xspace}

\begin{document}

\begin{frontmatter}

\title{Predicting \kovats Retention Indices Using Graph Neural Networks\tnoteref{t1}}

\tnotetext[t1]{Official contribution of the National Institute of Standards and Technology; not subject to copyright in the United States.}

\author[1]{Chen Qu}
\ead{chen.qu@nist.gov}

\author[1]{Barry I. Schneider}
\ead{barry.schneider@nist.gov}

\author[1]{Anthony J. Kearsley}
\ead{anthony.kearsley@nist.gov}

\author[1]{Walid Keyrouz}
\ead{walid.keyrouz@nist.gov}

\author[1]{Thomas C. Allison\corref{cor1}}
\ead{thomas.allison@nist.gov}

\cortext[cor1]{Corresponding author}

\address[1]{National Institute of Standards and Technology, 100 Bureau Drive, Gaithersburg, Maryland 20899, USA}

\begin{abstract}
The \kovats retention index is a dimensionless quantity that characterizes the rate at which a compound is processed through a gas chromatography column. This quantity is independent of many experimental variables and, as such, is considered a near-universal descriptor of retention time on a chromatography column. The \kovats retention indices of a large number of molecules have been determined experimentally.  The ``NIST 20: GC Method\slash Retention Index Library'' database has collected and, more importantly, curated retention indices of a subset of these compounds resulting in a highly valued reference database. The experimental data in the library form an ideal data set for training machine learning models for the prediction of retention indices of unknown compounds.  In this article, we describe the training of a graph neural network model to predict the \kovats retention index for compounds in the NIST library and compare this approach with previous work~\cite{2019Matyushin}. We predict the \kovats retention index with a mean unsigned error of 28 index units as compared to 44, the putative best result using a convolutional neural network~\cite{2019Matyushin}. The NIST library also incorporates an estimation scheme based on a group contribution approach that achieves a mean unsigned error of 114 compared to the experimental data. Our method uses the same input data source as the group contribution approach, making its application straightforward and convenient to apply to existing libraries. Our results convincingly demonstrate the predictive powers of systematic, data-driven approaches leveraging deep learning methodologies applied to chemical data and for the data in the NIST 20 library outperform previous models.
\end{abstract}


\end{frontmatter}


\section{Introduction}
Gas chromatography (GC) is an important analytical technique for the separation and identification of chemical compounds.  Frequently used in combination with mass spectrometry (GC/MS) as a means of enhancing the accuracy of identification of chemical compounds, a GC experiment consists of putting an unknown substance in a gaseous state into a carrier gas. This gas is passed through a chromatography column. Interactions with compounds used in the column determine the time elapsed before a compound elutes from (i.e., passes through) the column. The time a compound is retained in the column is indexed against elution times of known compound is called the retention index.

Work in the 1950s by Ervin \kovats~\cite{1958Kovats} demonstrated that the retention index could be made independent of many experimental factors such as column length, column diameter, and film thickness. This results in a dimensionless quantity known as the \kovats retention index, which is the quantity we consider in this article. (There are other retention indices not considered herein.)

In the context of identifying unknown compounds by searching a library of measured compounds, matching the retention index can significantly improve the confidence in results generated by library searching versus use of the mass spectrum alone.\cite{2003Eckel} Knowledge of retention indices thus has considerable value, and the ability to predict the retention index accurately constitutes a notable improvement to molecular libraries.

A wide variety of techniques have been employed to predict retention indices for various classes of compounds, many with considerable success. We briefly summarize the achievements of several studies below. Early studies using neural networks are characterized by small networks and small data sets of closely related compounds. Later studies consider larger data sets and more diverse sets of compounds. The present work was performed with the largest set of data currently available, significantly larger than those used in any other published investigation.

One of the earliest (1993) applications of neural networks to the prediction of the retention index is due to Bruchmann et al.~\cite{1993Bruchmann} who studied a set of terpenes. They discovered that their neural network was in good agreement with multilinear regression, but that the neural network did not generalize well to other classes of compounds (whereas multilinear regression performed better in this regard). In a 1999 publication, Pompe and Novi\v{c} used two artificial neural network approaches to predict retention indices for 381 compounds.\cite{1999Pompe} The input to their model consisted of 16 informational and topological descriptors calculated from molecular structures. This better model gave an error of 19.2 retention index units, outperforming a multilinear regression approach by 3.3.

Jalali-Heravi and Fatemi~\cite{2001JalaliHeravi} studied a set of 53 terpenes using a small artificial neural network and found agreement with experiment within \SI{2}{\percent}. Li et al.~\cite{2005Li} used topological indices and a neural network to estimate retention indices for 18 compounds. Using molecular descriptors such as boiling point and molecular weight as input to an artificial neural network, \v{S}krbi\'{c} and Onjia~\cite{2006Skrbic} successfully predicted Lee retention indices \cite{lee1979}. Their results agreed with experimental values within \SI{1.5}{\percent} over a set of 96 polycyclic aromatic hydrocarbon compounds.

In 2007, Stein et al.~\cite{2007Stein} published a group contribution method for the estimation of \kovats retention indices. Their data set consisted of more than \num{35000} molecules measured on different columns with different stationary phases and under differing thermal conditions. Their model made use of 84 groups, and they determined parameters for polar and nonpolar columns separately. The model predicted \kovats retention indices on polar and nonpolar columns with mean absolute errors (MAE) of 65 and 46 retention index units, respectively, corresponding to errors of \SI{3.9}{\percent} and \SI{3.2}{\percent}. The authors state that errors of this magnitude were too large to permit identification based on retention index alone, but could be used to reject certain identification errors in GC/MS library searches. In the present article, we are using a superset of the data that Stein et al. used to evaluate their model. The predictions of the group contribution model are used as a benchmark for the current neural network model.

The use of predicted retention indices to improve compound matching in GC/MS metabolomics experiments was taken up by Mihaleva et al.\cite{2009Mihaleva} A set of \num{22690} compounds were evaluated in their study. They constructed quantitative structure retention index models using multilinear regression (MLR) and support vector regression (SVR), finding that their SVR model performed better than either MLR or the group contribution method of Stein et al. for selected subsets of the data.\cite{2007Stein} They reported root mean square errors (RMSE) of 121 for the MLR method and 114 for the SVR method.

An extensive review article by Katritzky et al. in 2010 discusses artificial neural networks, support vector machines, and expert systems along with a variety of other prediction methods using quantitative structure property relationship (QSPR) methods.\cite{2010Katritzky} Their review includes discussion of the performance of a variety of QSPR methods for predicting retention indices and boiling points (which are strongly correlated with retention indices).

A trio of papers in 2011 explored the use of predicted retention index data to improve identification of unknown compounds in GC/MS experiments. Kumari et al.~\cite{2011Kumari} explored identification of compounds in metabolomics studies. The problem was studied more generally by Schymanski et al.~\cite{2011Schymanski} who used several predicted quantities, including the \kovats retention index from the model of Stein et al.~\cite{2007Stein}, to significantly reduce the number of possible matches from the large number of possible compounds based solely on the mass composition. Similarly, Zhang et al.~\cite{2011Zhang} used the method of Stein et al.~\cite{2007Stein} to construct distribution functions that were incorporated into a model to improve the matching of metabolites in rat plasma.

In the following year, Babushok and Andriamaharavo~\cite{2012Babushok,2007Babushok} demonstrated the use of the NIST retention index database to aid in identifying components found in an essential oil. Specifically, they found that differences between experimental and database values of retention indices provided substantial constraints on compound identification. In the same year, Schymanski et al.~\cite{2012Schymanski} published a followup study in which they tested their identification procedure on two complex environmental samples.

Yan et al.~\cite{2013Yan} employed multilinear regression, partial least squares, and support vector machine techniques to develop quantitative structure retention index relationships that were used to predict retention indices for compounds of plant essential oils. They found that support vector regression gave better performance while requiring a small number of variables.

Koo et al. and Zhang et. al~\cite{2014Koo,2011Zhang} reported an improved method for GC/MS compound identification as compared to their earlier work. The authors developed a method for reducing the standard deviation when combining retention index determinations from different column types and temperature programs. They demonstrated the efficacy of their approach in the determination of metabolites from mouse liver.

A recent paper by Anthony et al.~\cite{2018Anthony} examines the identification of compounds in GC/MS experiments,  highlighting the importance of the retention index in producing an improved result. The approach used in their paper is based on gas chromatographic (GC) separation of the unknown compounds followed by detection using a tandem vacuum ultraviolet-mass spectrometry (VUV-MS) instrument. Library searches for spectra matching in each of the three analytical dimensions (i.e. MS, GC, VUV) demonstrated the power of a multidimensional library search approach for reliable compound identification. 

In a 2018 paper, Zhokhov et al.~\cite{2018Zhokhov} discuss producing software that can predict retention indices with ``accuracy acceptable for analytical practice.'' This article contains an interesting discussion of the limitations and benefits of various methods for predicting retention indices including neural networks and group increment methods.

Recently, Matyushin et al.~\cite{2019Matyushin} applied a deep convolutional neural network methodology to the problem of predicting retention indices. The input to the model consisted of the simplified molecular-input line-entry system (SMILES) representation of the molecule.\cite{1988Weininger} The authors placed several limitations on the input, including a limit of 250 characters on the SMILES string and a restriction on the allowed atoms (we have similarly limited the types of atoms allowed in the present model for reasons described below). Information about stereo and geometric isomerism was discarded, and compounds that were considered identical had their retention index values averaged. The NIST 08 library was used, which contained fewer (experimental) determinations of the retention index than are available in the data set used in the present publication (NIST 20). On the test set employed by the authors, the model achieved a MAE of 33.2 retention index units (corresponding to a \SI{1.96}{\percent} MAE), whereas the model of Stein et al. yielded a 108.4 MAE.\cite{2019Matyushin, 2007Stein} As the methodology employed in this work is similar in some respects to our own work (in particular, both employ deep learning techniques), we will compare our model against the model of Matyushin et al.~\cite{2019Matyushin} extensively.

The present work makes use of the MatErials Graph Network (MEGNet) model developed by Chen et al.\cite{2019Chen} This model uses a graph convolutional neural network and has been used to produce state of the art performance in predicting properties of molecular and crystalline compounds using detailed structural information. (A survey article on graph neural networks by Wu et al.~\cite{2019Wu} gives an in-depth discussion of this methodology.) Graph convolutional networks represent a significant advance in machine learning that is particularly well suited for molecular problems.

The movement toward new machine learning architectures accelerated over the last few years, due to the existence of large data sets and much improved computational hardware and software, opening the opportunity for testing the limits of chemical property prediction. Early efforts used classical neural network architectures; much effort was dedicated to finding input representations that encoded increasing amounts of chemical information as well as addressing the problem of varying input vector length as molecule sizes (i.e., number of atoms) changed. A breakthrough publication by the Google Mind team used a deep learning architecture and demonstrated good performance for predicting total energies of molecules computed using density functional theory (DFT).\cite{2017Faber} A number of methodologies were developed in the following years that improved the accuracy and range of properties that could be predicted using machine learning models. These are important advances in this young and fast moving field of molecular and crystal property prediction using deep learning models.

However, these efforts are limited by the lack of large and diverse data sets. Most of the work to date has used the QM9 data set that contains \num{133885} molecules containing C, H, O, N, and/or F molecules with 9 or fewer C atoms.\cite{2009Blum,2012Rupp} These data sets rely on accurate geometric information obtained via optimization of the molecular structure using DFT with a modest basis set. Thus, the machine learning models are able to predict a relatively homogeneous (both chemically and methodologically) set of data. In particular, the error in the DFT energies and properties is systematic and thus amenable to learning as opposed to the various sources of noise present in experimental data (including outright errors). We thus temper our expectations and do not anticipate accuracy of 1 or 2 retention index units. Indeed, this conclusion is supported by the work of Matyushin et al.\cite{2019Matyushin}

Having set the stage for the importance and tractability of predicting the \kovats retention index for various molecules, we now describe our data set, model, and results with an emphasis on comparison with previous results and with a look to the future of machine learning in this area.

\section{Data and methods}

\subsection{Retention index data set}

The data for this study were obtained from NIST Standard Reference Database 1A: NIST\slash EPA\slash NIH Mass Spectral Library (NIST 20).\cite{nist20} The data contained in the mass spectral library are a subset of the data available in NIST 20: GC Method/Retention Index Library.\cite{2007Babushok} The mass spectral library set contains a 2D representation of the molecule (in Molfile format), the mass spectrum, a measured (i.e., experimental) value of the \kovats retention index, a predicted value of the \kovats retention index based on the model of Stein et al., an uncertainty on the predicted retention index value, and other data and metadata on the chemical compound. When multiple experimental determinations of the retention index are available, a consensus value is given along with the number of values and their range. Importantly, these data have been carefully curated by the personnel of the NIST Mass Spectrometry Data Center, yielding a large, high quality data set that is well suited for deep learning approaches.

The full NIST 20 library contains nearly \num{307000} compounds. For nearly 112,000 of these molecules, an experimental \kovats retention index is available. The library contains data from 3 different column types. Among these, the largest collection of experimentally determined \kovats retention indices is for the ``semi-standard non-polar'' column type. Sets of experimental values for standard polar and non-polar columns are an order of magnitude smaller. Thus, we have restricted the present study to data from experiments using the semi-standard non-polar column type, yielding a data set with \num{104896} molecules. These data may be either single experimental values or a consensus value produced from more than one measurement.

In order to ensure that the final model contained sufficient examples of each atom type, any molecule containing an atom that appeared in less than 50 molecules in the data set was removed. Therefore, only molecules containing (in order of decreasing frequency) C, O, N, Si, F, Cl, S, Br, H, I, and P atoms remained. The atom appearing in the fewest number of molecules, P, was present in 994 molecules in the data set, ensuring good representation during training. (H atoms are not represented in the input file except where required to adequately describe a molecular structure, hence the low frequency of H atoms in the preceding list.) We next examined histograms of molecular mass and retention index vales, seeking to ensure a reasonable density of data at the extreme values of these two quantities to avoid problems during model training. Molecules with mass less than 50~amu and more than 850~amu were removed from the set. Similarly, molecules with \kovats retention indices less than 300 and more than \num{4100} were omitted. (This procedure removed 246 molecules from the data set.)

The removal of molecules on the basis of mass and \kovats retention index value give an indication of where the present model should work best. Although machine learning methods can sometimes extrapolate effectively, this is not something the data permits us to test effectively, especially given the small number of compounds removed. In the case of the mass limit, predictions of derivatized compounds whose mass may easily exceed 850~amu should be used with appropriate care.

The data set used for training, validation, and testing (after the removals described above) consisted of \num{104650} molecules. A histogram of the retention index data is shown in Figure~\ref{fig:hist-data}. As can be seen in the figure, the data are nearly Gaussian distributed ($\mu=\num{2142}$, $\sigma=\num{642}$). These data will be used for training and testing the model after normalization via computing the z-score,
\begin{equation}
y'=\frac{y-\mu_y}{\sigma_y}
\end{equation}
where $y'$ is the value of the normalized experimental \kovats retention index, $y$ is the corresponding unnormalized experimental value, $\mu_y$ is the mean of $y$, and $\sigma_y$ is the standard deviation of $y$.

\begin{figure}
\centering
[Approximate location of figure \ref{fig:hist-data}]
\caption{Histogram of the full \kovats retention index data set used for training, validation, and testing of the model. The red line plots a Gaussian distribution with the same mean and standard deviation as the data set ($\mu=\num{2141.6}$, $\sigma=\num{641.9}$).}
\label{fig:hist-data}
\end{figure}

The features extracted from the NIST 20 data set are the atomic identities and the list of chemical bonds (from the MolFile representation of the molecule). Our goal in this work is to produce a machine learning model capable of giving accurate predictions of the \kovats retention index with a minimum of chemical information needed as input. We deliberately restrict the model input to information on a molecule that can be obtained from a familiar 2D sketch of the molecule or obtained from a library of molecular structures (as we have done in this work). It is hoped that by developing a model that relies only on {\it simple} structural information, the barrier to using the methodology for making predictions is as low as possible.

\subsection{Graph Neural Network Methodology}

As mentioned previously, we make use of the MatErial Graph Network (MEGNet) methodology developed by Chen et al.\cite{2019Chen} The MEGNet model incorporates a number of features that make it a good choice for machine learning of molecular properties. In particular, a graph neural network is employed.\cite{2018Battaglia} The graph used by the machine learning model corresponds directly to the molecular structure. The nodes/vertices in the graph correspond to the locations of the atoms, and the edges correspond to atomic pairs (there is an edge between any pair of atoms even if they are not formally chemically bonded). Thus, the chemical graph model may be directly mapped onto the graph network model. Properties of atoms and bonds may be assigned to the nodes and edges in the graph model. The MEGNet framework also allows global (i.e. not associated with a single atom or chemical bond) properties of a molecule to be incorporated into the model.

The MEGNet model is thoroughly described in the literature and online, where the source code is also available.\cite{2019Chen, megnet} We will only repeat the salient features of the model so that it may be reproduced by others. MEGNet version 0.3.5 is used with TensorFlow version 1.14.1 and the Keras API. MEGNet encodes input data in 3 distinct vectors (called ``attributes''): state attributes, edge attributes, and node attributes. State attributes are global properties of the molecule (e.g., numbers of atoms, molecular mass). Edge attributes are features of atom pairs (e.g., bond order, member of ring). Node attributes are features of the atoms themselves (e.g., atomic number, hybridization). Each of these attributes are used to update the other attributes and are themselves updated during model fitting.

Our MEGNet model incorporates 3 atom features (encoded as 18 one-hot variables), 3 edge features (encoded as 7 one-hot variables), and 3 global features. The atom features are the atomic number (11 one-hot variables representing C, H, O, N, F, Si, P, S, Cl, Br, and I), the hybridization of the atom (6 one-hot variables, $s$, $sp$, $sp^2$, $sp^3$, $sp^3d$, and $sp^3d^2$), and the formal charge (1 integer variable) on the atom. The hybridization is calculated from the information in the MolFile using RDKit. The edge features are the bond type (i.e., no bond, single, double, triple, or aromatic, 5 one-hot variables), whether the atoms were in the same ring (a single one-hot variable), and a graph distance (1 integer variable). The graph distance is calculated as the smallest number of edges that have chemical bonds between the atoms in the pair (recall that an edge here means any pair of two atoms, not just those that are chemically bonded). We do not consider pairs with a graph distance higher than 5 to save computational resources as discussed below. Global features are the number of heavy (non-hydrogen) atoms in the molecule, the molecular mass divided by the number of heavy atoms, and the number of chemical bonds divided by the number of heavy atoms.

Our MEGNet model was used with 3 MEGNet blocks. These blocks are composed of two layers of densely-connected multi-layer perceptrons (MLP) and a graph neural network layer in which each of the attributes is successively updated. The dense layers used 128 and 64 units, respectively. The MEGNet block steps are followed by a `Set2Set' layer~\cite{2015Vinyals} in which the output of the bond and atom attributes are mapped to the appropriate vector quantities. This is followed by a concatenation step and two densely-connected MLPs (64 and 32 units).

We have modified the MEGNet source code to limit the depth of atom pair properties in order to save computer memory, which scales roughly as the square of the number of atoms times the number of features. With up to 151 atoms in a molecule, the CPU memory requirements ran into the 100s of GB during model development. (GPU memory was not a bottleneck when training the model, however, due to batching.) We also implemented a custom molecule class that relied on features generated using RDKit.\cite{rdkit}

Training was carried out for up to \num{2000} epochs, with early stopping employed with restoration of the best weights if the value of the loss function for the validation set did not improve for 150 epochs. The average number of epochs in training was approximately \num{1400}. A batch size of 32 was used during training. The Adam optimizer was used in fitting the model, with a learning rate of $2\times 10^{-4}$. The MAE was used as the loss function instead of the usual mean squared error (MSE) commonly used in training neural network models. The rectified linear unit (ReLU) activation function was used in the MLP layers. Interestingly, the MEGNet paper describes use of the softmax activation function, but we found better results were produced using ReLU activation. The output layer consisted of a single MLP unit with linear activation. The experimental values of the \kovats retention index as taken from the NIST 20 library were used as target values. These were normalized as described previously.

Prior to training, the data set was split into three sets: training, validation, and testing in an 8:1:1 ratio. This was accomplished by randomly dividing the full data set into 10 blocks and then using 8 blocks for training with the remaining blocks used for validation and testing. Due to the large size of the data set, our testing set includes more than \num{10000} retention indices. This division of data was convenient for the 10-fold cross-validation procedure we employed to test the reliability of our model.

\section{Results and discussion}

The model described above was produced via a trial and error process, considering a large number of features and testing the sensitivity of the model to including/removing various features. For example, atom features for aromaticity, ring atoms, and ring size did not significantly improve the model. This is likely because these features contain information contained in bond features. Similarly, for bond features, we attempted to use 3D geometry information from conversion of 2D structures and force field minimization, but abandoned this approach because we did not see significant improvement and did not want to unnecessarily complicate the model given our particular use case.

We experimented with several different loss functions in an attempt to minimize both the final MAE and the standard deviation of the absolute error of the predicted retention indices versus the experiment. Initial runs made use of the MSE as is common in training machine learning models. We found that switching to the MAE as the loss function produced results with a lower MAE than training with the MSE loss function, while leaving the value of the standard deviation of the error approximately constant. We tried weighted and unweighted combinations of MAE and MSE as well as combinations of the $\ell_1$, $\ell_2$, and $\ell_\infty$ norms. None of these produced results that were superior to the MAE in reducing the mean and standard deviation of the error, so the MAE was used as the loss function for all subsequent model training.

Our model training makes use of training, validation, and testing sets. As there is some inconsistent usage of these terms in the literature, we describe them here. The training set is used in the training of the model. For this work, we used \SI{80}{\percent} of the data for this purpose. The validation set consisted of \SI{10}{\percent} of the data and was used to evaluate model performance during training. Use of a validation set can be a good way to produce more robust machine learning models, reducing the mean and standard deviation of the error in the final results. Use of the validation set made a small improvement to our final results. Finally, the testing set was composed of the remaining \SI{10}{\percent} of the data and was used for evaluating the performance of the model using various performance metrics described below. As the training and validation sets are used in model fitting, they do not provide an unbiased estimate of the model error. However, testing on a fraction of the data can lead to models that are biased. To overcome this problem, $k$-fold cross validation may be employed.

Once we arrived at a model with acceptable performance, we used a 10-fold cross-validation scheme to test the reliability of the model. Cross validation has been shown to provide a robust test of machine learning models. To carry out $k$-fold cross validation, one divides the data set into $k$ sets (or ``folds'') and then trains the model $k$ times, each time with $k-1$ folds in the training set and the remaining fold in the testing set. Using this procedure, every element in the data set is used for testing the model.

The results of the 10-fold cross-validation procedure are given in Table~\ref{table:crossval}. Our model produces an MAE of 28.09 index units and an RMSE of 58.43 index units. This represents a modest improvement versus the results of Matyushin et al.\cite{2019Matyushin} who reported values of 33.2 and 63.0 for the MAE and RMSE, respectively. However, the present results were obtained with a set of data more than 3 times the size of that used by Matysuhin et al. To test the two neural network models more fully, we implemented the convolutional neural network model of Matyushin et al. in Python using Tensorflow and applied it to the new NIST 20 dataset. Our implementation of that model achieved an MAE of 44.4 and an RMSE of 74.6 index units in a 10-fold cross-validation test, both of which are indeed larger than those of our graph network model. Our results represent a significant improvement over the model of Stein et al.\cite{2007Stein} with an MAE of 114.3 and an RMSE of 166.5 (standard deviation of 121.1). These larger errors are likely due partially to deficiencies in the group increment methodology and partially due to the use of the less powerful ``classical'' machine learning method (linear least squares) for model optimization.

\begin{table}
\centering
\begin{tabular}{@{}l c c@{}}
\toprule
Set        &      MAE         &       RMSE       \\
\midrule
Training   &  9.38 $\pm$ 0.86 & 18.27 $\pm$ 1.87 \\
Validation & 27.84 $\pm$ 0.67 & 57.77 $\pm$ 1.97 \\
Testing    & 28.09 $\pm$ 0.72 & 58.43 $\pm$ 1.93 \\
\bottomrule
\end{tabular}
\caption{Statistics of the 10-fold cross validation procedure. The mean value and standard deviation of the mean absolute error (MAE) and the root mean square error (RMSE) over 10 runs is given for each of the 3 sets used.}
\label{table:crossval}
\end{table}

In Table~\ref{table:crossval}, it is clear that there is overfitting present in the model as evidenced by the increase in the MAE of 19 index units and in the RMSE of 40 index units. This is also evident in the plot of the training history presented in Figure~\ref{fig:training-history}. This figure also shows that the MSE (not used as a loss function during training) follows a similar pattern as the MAE, another indication of good fitting. The plots of MAE and MSE show a rapid decrease in the error metric in the first 50 epochs followed by a slow drop to a minimal value. There does not seem to be any significant improvement in the performance of the model past 400 epochs. This behavior does not negatively affect our model because we recover the best model weights when early stopping is employed as is the case in the figure. The similar values of the validation and testing set errors are an indication that the validation set is effective. Despite overfitting, the MAE and RMSE of the testing set are acceptable and are indicative of a good model for prediction of the \kovats retention index. The sample standard deviations of both error metrics over the 10-fold cross validation procedure indicate that the model fitting is robust. The sample standard deviations for the training set are likely larger due to the larger amount of data in that set. However, the smaller sample standard deviation of the error metrics in the testing set is a strong indication of good model performance.

\begin{figure}
\centering
[Approximate location of figure \ref{fig:training-history}]
\caption{Plot of the training and validation loss function (MAE, upper panel) versus the training epoch. It can be clearly seen that overfitting sets in around 50--100 epochs but is stable thereafter. The same plot for the mean squared error (MSE), not used as the loss function during training, is given in the lower panel.}
\label{fig:training-history}
\end{figure}

In order to further characterize the performance of the model, the results from a single run in the cross-validation set are now analyzed in detail. These results are presented in Table~\ref{table:stats}. This run has an MAE nearly identical to the mean produced by the 10-fold cross validation and thus serves as a good representative of the model. There are several interesting details to notice in the table. The standard deviation of the absolute error (i.e., the RMSE) is consistently larger than the standard deviation of the signed error. The mean (absolute) percent error (MPE) is \SI{1.49}{\percent} for the testing set with a standard deviation of \SI{2.84}{\percent}. This compares favorably to MPE values of \SI{1.96}{\percent} from the model of Matyushin et al.~\cite{2019Matyushin} and \SI{5.4}{\percent} for the model of Stein et al.\cite{2007Stein} The largest errors produced by the model are significant. The source of these deviations is not clear and may be due to an inadequacy in our model or due to an error in the source data. (We are currently investigating whether large prediction errors in our model correspond to errors in the source experimental data.)

\begin{table}
\centering
\begin{tabular}{@{}l r r r@{}}
\toprule
Quantity                  & Training    & Validation  & Testing \\
\midrule
$n$                       & \num{83719} & \num{10464} & \num{10464} \\
mean $\epsilon$           & \num{-1.06} & \num{-1.23} & \num{-0.47} \\
RMSE                      & \num{20.69} & \num{54.61} & \num{57.90} \\
min $\epsilon$            & \num{-1163} & \num{-979}  & \num{-832}  \\
max $\epsilon$            & \num{1013}  & \num{586}   & \num{1175}  \\
median $|\epsilon|$       & \num{6.60}  & \num{12.20} & \num{12.50} \\
mean $|\epsilon|$         & \num{10.20} & \num{27.69} & \num{28.67} \\
$s(|\epsilon|)$           & \num{18.03} & \num{47.09} & \num{50.30} \\
median $|\%\epsilon|$     & \num{0.32}  & \num{0.60}  & \num{0.62}  \\
mean $|\%\epsilon|$       & \num{0.52}  & \num{1.42}  & \num{1.49}  \\
$s(|\%\epsilon|)$         & \num{1.00}  & \num{2.56}  & \num{2.84}  \\
max $|\%\epsilon|$        & \num{73.60} & \num{49.13} & \num{66.74} \\
\bottomrule 
\end{tabular}
\caption{Statistics describing the graph convolutional network model performance in predicting the \kovats retention index (RI) for the training, validation, and testing sets. The error is calculated as $\epsilon=\mathrm{RI_{experiment}}-\mathrm{RI_{predicted}}$. The sample standard deviation, $s$, is calculated for the signed and unsigned errors.}
\label{table:stats}
\end{table}

The model performance is further described using two figures. In Figure~\ref{fig:hist-error}, the absolute errors for all three sets (training, validation, and testing) are shown in a histogram plot. This figure clearly shows that most errors are small and the number of compounds with larger errors decreases monotonically. In Figure~\ref{fig:compare}, the predicted data are plotted versus the experimental values taken from the NIST 20 database. The red line shows 2 standard deviations in the MAE for each of the three sets.

\begin{figure}
\centering
[Approximate location of figure \ref{fig:hist-error}]
\caption{Histogram of mean absolute errors in predicting the \kovats retention index (experimental $-$ predicted) in the training, validation, and testing data sets.}
\label{fig:hist-error}
\end{figure}

\begin{figure}
\centering
[Approximate location of figure \ref{fig:compare}]
\caption{Comparison of predicted to actual values of the \kovats retention index. The red line indicates perfect agreement, with the width of the line equal to 2 standard deviations of the prediction error.}
\label{fig:compare}
\end{figure}

In order to further characterize the performance of the model presented in this work, we performed comparisons with the group contribution model of Stein et al. \cite{2007Stein} and the convolutional neural network (CNN) model of Matyushin et al. \cite{2019Matyushin}. The models of Matyushin et al.\cite {2019Matyushin} was reimplemented by us using TensorFlow \cite{tensorflow} and RDKit \cite{rdkit}. To fully characterize this model, we performed 10-fold cross validation using the same learning rate as was used for the training of our model.

Table \ref{table:comparison1} presents mean absolute errors and the sample standard deviation of the absolute mean errors for the three prediction models indicated for the indicated chemical functionalities. The results are ordered by the MAE of the present model, and the statistics of all compounds are included for comparison purposes. In all of the cases presented, the model of Stein et al. \cite{2007Stein} exhibits the largest error statistics of the three models. This is consistent with the conclusions drawn by Matyushin et al. \cite{2019Matyushin} and from the present work. The Stein model works particularly well for hydrocarbons and exhibits the largest errors for O heterocycles. Strangely, the estimate of the Stein model is given in NIST20 for only 35 of the 994 compounds containing P. Using the full set of P compounds improves the statistics as shown in the second entry labelled ``contains P'' in the table for both machine learning models. However, in both cases the sample standard deviation is largely unchanged even though the MAE decreases appreciably.

The results from Table \ref{table:comparison1} allow us to explore the performance of the CNN model \cite{2019Matyushin} versus the present model. The MAE for the CNN model is 5.50 to 14.85 retention index units higher than the results from our graph neural network (GNN) model. When expressed as a percentage, the MAE values produced by the CNN are 13\% to 51\% higher than the results from the GNN. When the same comparison is done for the sample standard deviation, the CNN performs better with values of 0.56 to 8.85 retention index units higher than the GNN or from nearly equal up to 19\% larger than the GNN. (Note the we have used the updated statistics for P containing molecules presented in the paragraph above for this analysis.) We note that the performance of the CNN model is generally very good as might be expected from the analysis performed by Matyushin et al. \cite{2019Matyushin}. We suggest that the superior results for our model are likely due to the power of the graph neural network, particularly when applied to chemical problems, and to the larger size of our model. Finally, we note that the input data needed for either model is the same -- a molecular representation that captures connectivity -- and that conversion between the input formats (SMILES for the CNN and MolFile for the GNN) is wasily accomplished by a number of readily-available software libraries.

The largest errors produced by the two neural network models come from molecules with hydroxyl groups, i.e. alcohols and carboxylic acids. Performance on heterocycles, ketones, and molecules that contain S also have larger MAE values as well as larger sample standard deviations. While we are not aware of any systematic bias in either model that would favor one class of compound over another, these results suggest that there may be room for improvement by explicitly considering these factors. However, this could also be a symptom of data with larger uncertainties.

\begin{table}
\centering
\begin{tabular}{ l r r@{.}l r@{.}l r@{.}l r@{.}l r@{.}l r@{.}l }
\toprule
Molecule & & \multicolumn{4}{c}{Stein} & \multicolumn{4}{c}{Matyushin} & \multicolumn{4}{c}{present work} \\
type & $N$ & \multicolumn{2}{c}{$\mu$} & \multicolumn{2}{c}{$s$} & \multicolumn{2}{c}{$\mu$} & \multicolumn{2}{c}{$s$} & \multicolumn{2}{c}{$\mu$} & \multicolumn{2}{c}{$s$} \\
\midrule
ether           &  55755 & 112 & 60 & 120 & 21 &  34 & 66 &  52 & 16 &  22 & 99 &  43 & 87 \\
amide           &  24033 & 119 & 26 & 112 & 20 &  38 & 75 &  56 & 73 &  25 & 96 &  47 & 88 \\
contains O      &  92136 & 114 & 36 & 119 & 88 &  38 & 03 &  56 & 90 &  26 & 65 &  49 & 64 \\
\\[-3ex]
{\it all compounds} & 102761 & 113 & 84 & 119 & 44 &  38 & 97 &  57 & 08 &  27 & 74 &  50 & 00 \\
\\[-3ex]
hydrocarbon     &   2010 &  61 & 67 &  94 & 10 &  33 & 27 &  49 & 57 &  27 & 77 &  46 & 75 \\
aromatic        &  70501 & 121 & 93 & 119 & 82 &  42 & 48 &  59 & 36 &  29 & 68 &  50 & 95 \\
has a ring      &  79091 & 125 & 44 & 125 & 90 &  43 & 37 &  60 & 77 &  30 & 80 &  53 & 19 \\
contains N      &  51536 & 124 & 71 & 119 & 04 &  46 & 78 &  64 & 90 &  33 & 51 &  56 & 98 \\
aldehyde        &   1176 & 114 & 23 & 121 & 06 &  45 & 89 &  55 & 80 &  38 & 66 &  49 & 55 \\
contains S      &   9707 & 127 & 60 & 120 & 98 &  52 & 85 &  71 & 25 &  39 & 87 &  64 & 53 \\
N heterocycle   &  22382 & 155 & 50 & 138 & 27 &  61 & 12 &  75 & 80 &  46 & 49 &  67 & 99 \\
ketone          &   5611 & 168 & 16 & 172 & 95 &  60 & 98 &  80 & 86 &  46 & 94 &  72 & 89 \\
O heterocycle   &   9488 & 182 & 43 & 173 & 88 &  63 & 13 &  73 & 99 &  48 & 28 &  66 & 49 \\
alcohol         &   7102 & 141 & 22 & 153 & 86 &  64 & 01 &  80 & 56 &  52 & 26 &  75 & 42 \\
carboxylic acid &   1444 & 101 & 78 & 123 & 98 &  63 & 44 &  95 & 51 &  56 & 09 &  94 & 95 \\
contains P      &     35 & 131 & 00 & 172 & 38 &  66 & 18 &  79 & 00 &  66 & 31 &  76 & 85 \\
contains P      &    994 & \multicolumn{2}{c}{} & \multicolumn{2}{c}{} & 39 & 64 &  79 & 09 &  31 & 70 &  73 & 73 \\
\bottomrule 
\end{tabular}
\caption{Evaluation of the performance of the model of Stein et al. \cite{2007Stein}, the model of Matyushin et al. \cite{2019Matyushin}, and the model described in this article for selected chemical functionalities for the (unitless) \kovats retention index. The column labelled $N$ indicate the number of data used in computing the mean ($\mu$) and sample standard deviation ($s$) of the absolute error, $|\epsilon|=|\mathrm{RI_{experiment}}-\mathrm{RI_{predicted}}|$.}
\label{table:comparison1}
\end{table}

The performance of the three models is further demonstrated by comparing error statistics for collections of molecules. This analysis is presented in Table \ref{table:comparison2}. The collections of molecules were gathered from various literature sources as follows. The list of ``flavors and fragrances'' were taken from the work of Rojas et al. \cite{2015Rojas1}. The list of ``fragrance-like'' molecules was taken from another publication by Rojas et al. \cite{2015Rojas2}. Another list of ``flavors'' was taken from the work of Yan et al. \cite{2013Yan}. A list of ``essential oils'' was taken from the work of Babushok et al. \cite{2011Babushok}. Finally, the list of ``terpenes'' was taken from a paper by Jalali-Heravi et al. \cite{2001JalaliHeravi}. For the first three sets of molecules, SMILES strings were provided and were used for matching compounds in our data set. SMILES string for our data set were generated using RDKit \cite{rdkit}. For the essential oil compounds, matches were made using CAS registry numbers.

The data in Table \ref{table:comparison2} show marked improvement in the error statistics for all three estimation schemes. The improvement is most pronounced for the model of Stein et al. \cite{2007Stein}. Nevertheless, the pattern that emerged in the previous analysis remain intact. The CNN model of Matyushin et al. \cite{2019Matyushin} exhibits errors approximately half of those from the Stein model, with the GNN model of the present work giving further improvement over the CNN results.

\begin{table}
\centering
\begin{tabular}{ l r r@{.}l r@{.}l r@{.}l r@{.}l r@{.}l r@{.}l }
\toprule
Molecule & & \multicolumn{4}{c}{Stein} & \multicolumn{4}{c}{Matyushin} & \multicolumn{4}{c}{present work} \\
type & $N$ & \multicolumn{2}{c}{$\mu$} & \multicolumn{2}{c}{$s$} & \multicolumn{2}{c}{$\mu$} & \multicolumn{2}{c}{$s$} & \multicolumn{2}{c}{$\mu$} & \multicolumn{2}{c}{$s$} \\
\midrule
flavors and \\
fragrances & 675 & 37 & 94 & 43 & 59 & 18 & 96 & 24 & 40 & 16 & 30 & 17 & 93 \\
fragrance- \\
like       & 662 & 36 & 81 & 39 & 10 & 18 & 46 & 21 & 48 & 16 & 13 & 17 & 87 \\
flavors    & 354 & 34 & 32 & 31 & 60 & 19 & 47 & 22 & 70 & 15 & 60 & 17 & 90 \\
essential \\
oils       & 473 & 41 & 04 & 40 & 86 & 21 & 94 & 22 & 87 & 19 & 09 & 20 & 37 \\
terpenes   &  39 & 31 & 82 & 20 & 99 & 15 & 35 & 13 & 91 & 11 & 89 & 10 & 33 \\
\bottomrule
\end{tabular}
\caption{Evaluation of the performance of the model of Stein et al.\cite{2007Stein}, the model of Matyushin et al.\cite{2019Matyushin}, and the model described in this article for selected molecule types. The column labelled $N$ indicates the number of data used in computing the mean ($\mu$) and sample standard deviation ($s$) of the absolute error, $|\epsilon|=|\mathrm{RI_{experiment}}-\mathrm{RI_{predicted}}|$. See text for discussion.}
\label{table:comparison2}
\end{table}

The goal of the model presented in this article is to predict the \kovats retention index with good accuracy. The best model is characterized by small values of both the MAE and the standard deviation of the error. Our experience in working with this data set leads us to believe that further improvement may not be possible due to the uncertainty present in the experimental data. However, it is certainly possible that a more powerful machine learning model may lead to even better results. In the context of the present work, this might be accomplished by using additional atom, bond, and global features. We note that, consistent with the approach used by some of the best contemporary machine learning models, we could generate accurate molecular geometries using quantum chemistry techniques. However, this approach requires a significant amount of computer time and would violate our original goal of creating a model that was a drop-in replacement for models used in databases.

Graph convolutional networks have proven to be a very powerful approach for producing models with state-of-the-art accuracy as exemplified by the MEGNet methodology.~\cite{2019Chen} For example, the MEGNet model was used to fit 13 properties of the QM9 data set~\cite{2009Blum, 2012Rupp} with accuracy close to or exceeding SchNet\cite{2018Schutt} and the neural message passing methodology previously applied to the same data set.\cite{2017Gilmer, 2018Jorgensen} The results produced in the present study do not meet the incredibly low mean absolute error targets enjoyed by these other studies. There are several reasons that this is the case. First, the other studies were performed on the QM9 data set.\cite{2012Ruddikeit, 2014Ramakrishnan} This data set consists of \num{133885} species with up to nine heavy atoms (C, O, N, and F), with properties computed using DFT at the B3LYP/6-31G(2df,p) level of theory. As such, it comprises a set created with consistent methodology and presumably has consistent error behavior as well. In contrast, the present work has approximately half of the data values, but explores a much more varied chemical space. For example, our data set has twice as many heavy atoms (i.e., atoms other than H), and contains molecules with up to 73 heavy atoms. Moreover, studies based on the QM9 data set use coordinates from the optimized geometries of the molecules to compute bond lengths to which chemical properties are sensitively related. In contrast, we do not use bond length, preferring to use only an indication of whether two atoms are bonded due to the nature of our intended target. Most importantly, our target data set consists of experimental values which are necessarily ``noisy'' and likely to include errors. So, while achieving as small an MAE as possible is desirable, we recognize that it is unlikely that we will be able to achieve MAEs of a few retention index units.

We did attempt to construct a model using SchNet, but were not able to achieve a model with acceptable accuracy. As noted above, SchNet is capable of state of the art accuracy on molecular properties. In this case, we believe the requirement that properties are calculated as a sum over atomic contributions was not appropriate for predicting the retention index, a quantity that is determined by dynamical interactions of a molecule with other molecules in the chromatography column rather than internal contributions. Given that the group increment model makes a similar assumption at the level of chemical groups rather than atoms, we suggest that the group increment model may be similarly deficient.

We also attempted to construct a traditional neural network (multilayer perceptron) model by using histograms of inter-atomic distances and angles such as the methodology proposed by Collins et al.\cite{2018Collins} As with other methods, we did not find that this method led to accurate results, though we do not preclude the possibility that a successful model could be constructed in this manner.

Finally, we note that there is significant value in predictive models for the \kovats retention index for mass spectral library searching. The improved performance afforded by the current model should lead to corresponding improvements in library searches, smaller lists of possible matches with fewer false positives.

The NIST library of retention index values is constantly growing and being refined. We anticipate that future releases of the library will permit even more accurate models that cover an even greater range of chemical functionalities. The value of such libraries, carefully curated by experts, cannot be overstated, particularly when used in machine learning applications.

\section{Conclusion}
In this article, we have demonstrated the application of a graph neural network to the problem of predicting an experimentally-derived set of \kovats retention indices. Our model requires only the information that can be readily obtained from a 2D representation of the molecule (i.e., a traditional drawing of a molecule). We have trained our model using a diverse set of more than \num{100000} molecules, using \SI{90}{\percent} of the data set in training and validation and using the remaining \SI{10}{\percent} of the data set for testing the performance of the model. We find that we are able to predict the \kovats retention index with an MAE of 28 (with a corresponding RMSE of 58) versus the experimental data. The performance of the current model is very similar to the one reported by Matyushin et al.\cite{2019Matyushin}. However, the current model used a data set three times larger than the one used by Matyushin et al. and our experience shows that larger data sets tend to produce a larger MAE. Reparameterizing the model by Matyushin et al. using the latest NIST 20 data set yields slightly larger MAE and RMSE compared to our model. Notably, both our model and that of Matyushin et al. perform significantly better than the widely used model of Stein et al., based on group increment values which has an MAE of 114 and a standard deviation of 167 over the same data set.~\cite{2007Stein} Both prediction methods use the same input, namely a 2D representation of the molecule as can be easily created using a variety of chemical drawing programs. This approach has the benefit that estimates of the \kovats retention index may be rapidly generated without the need for more detailed structural information that might be computed using quantum chemistry programs at a cost of considerably more time. Conversely, the lack of detailed structural information that is common in other machine learning applications may ultimately limit the performance of the model used in this work.

\section*{Acknowledgement}
The authors are grateful to Dr. William Wallace for providing the NIST 20 data set and for helpful discussions pertaining to the content and usage of the data.

\bibliography{predictingRI.bib}

\end{document}